\documentclass[useAMS,usenatbib]{mn2e}

\usepackage[psamsfonts]{amssymb}
\usepackage[dvips]{graphicx}
\usepackage{graphicx}
\usepackage{amsmath,alltt}                                                                                                                          
\usepackage{multirow}
\usepackage{rotating}
\usepackage{lscape}
\usepackage{tablefootnote}
\usepackage{hyperref}
\title[Kicked BHs in rotating star clusters]{The evolution of kicked stellar-mass black holes in star cluster environments II. Rotating star clusters}
\author[Webb et al.]{Jeremy J. Webb$^1$\thanks{E-mail: webb@astro.utoronto.ca (JW)}, Nathan W. C. Leigh$^{2,3,4}$, Roberto Serrano$^{2,3}$, Jillian Bellovary$^{3,5}$,
\newauthor K. E. Saavik Ford$^{3,6,7}$, Barry McKernan$^{3,6,7}$,Mario Spera$^{8,9,10}$ \& Alessandro A. Trani$^{11}$ \\
$^1$ Department of Astronomy and Astrophysics, University of Toronto, 50 St. George Street, Toronto ON M5S 3H4, Canada \\
$^{2}$Department of Physics and Astronomy, Stony Brook University, Stony Brook, NY 11794-3800, USA\\
$^{3}$Department of Astrophysics, American Museum of Natural History, Central Park West and 79th Street, New York, NY 10024 \\
$^{4}$Departamento de Astronom\'a, Facultad de Ciencias F\'sicas y Matem\'aticas, Universidad de Concepci\'on, Concepci\'on, Chile \\
$^{5}$ Department of Physics, Queensborough Community College, Bayside, NY 11364 \\
$^{6}$ Department of Science, Borough of Manhattan Community College, City University of New York, New York, NY 10007 \\
$^{7}$ Physics Program, The Graduate Center, CUNY, New York, NY 10016 \\
$^{8}$ Dipartimento di Fisica e Astronomia ‘G. Galilei’, University of Padova, Vicolo dell’Osservatorio 3, I–35122, Padova, Italy \\
$^{9}$ Department of Physics and Astronomy, Northwestern University, Evanston, IL 60208, USA \\
$^{10}$ Center for Interdisciplinary Exploration and Research in Astrophysics (CIERA), Evanston, IL 60208, USA \\
$^{11}$Department of Astronomy, Graduate School of Science, The University of Tokyo, 7-3-1 Hongo, Bunkyo-ku, Tokyo, 113-0033, Japan}
\begin{document}

\pagerange{\pageref{firstpage}--\pageref{lastpage}} \pubyear{2018}

\maketitle

\label{firstpage}

\begin{abstract}
In this paper, we continue our study on the evolution of black holes (BHs) that receive velocity kicks at the origin of their host star cluster potential.  We now focus on BHs in rotating clusters that receive a range of kick velocities in different directions with respect to the rotation axis.  We perform $N$-body simulations to calculate the trajectories of the kicked BHs and develop an analytic framework to study their motion as a function of the host cluster and the kick itself. Our simulations indicate that for a BH that is kicked outside of the cluster's core, as its orbit decays in a rotating cluster the BH will quickly gain angular momentum as it interacts with stars with high rotational frequencies. Once the BH decays to the point where its orbital frequency equals that of local stars, its orbit will be circular and dynamical friction becomes ineffective since local stars will have low relative velocities. After circularization, the BH's orbit decays on a longer timescale than if the host cluster was not rotating. Hence BHs in rotating clusters will have longer orbital decay times.  The timescale for orbit circularization depends strongly on the cluster's rotation rate and the initial kick velocity, with kicked BHs in slowly rotating clusters being able to decay into the core before circularization occurs. The implication of the circularization phase is that the probability of a BH undergoing a tidal capture event increases, possibly aiding in the formation of binaries and high-mass BHs.

\end{abstract}


\section{Introduction} \label{intro}

As star clusters evolve through cosmic time, black holes (BHs) inevitably form by stellar evolution. Upon formation, newly formed BHs may receive natal kicks up to $10^2$ km/s \citep{repetto12}. A cluster's BH population will quickly sink through the ``sea of stars" toward the cluster centre due to stellar dynamical friction (sDF) \citep{Chandrasekhar43}. sDF acts to decelerate the most massive objects in a population the fastest, leading to the rapid mass segregation of BHs and massive stars. Once the cluster fully relaxes, all of the BHs will be concentrated at the minimum of the cluster's potential while lighter stars will tend to orbit farther away from the core. In the cores of nuclear and possibly globular clusters, massive BHs (MBHs; M$_{\rm BH} \gtrsim$ 100 M$_{\odot}$) will typically reside surrounded by the highest stellar densities known to astronomical observations. 

Once in the core, BHs have several observational implications for their host cluster. The complex dynamical interplay between BHs and cluster stars ultimately shapes the evolution of clusters and the fate of any BHs residing therein. For example, BHs affect the evolution of star clusters by impacting their mass segregation rates and core collapse times \citep{Trenti10, Lutzgendorf13, Webb16, Alessandrini16}. For individual BHs, the relevant timescales over which kicked BHs decay to the cluster's core, encounter other stars and BHs to form binaries, and grow in mass through binary mergers and the consumption of stars are all dependent on both the properties of the BHs themselves (including the range of kick velocities that they can receive) and the host star cluster. 

The stellar distribution around a MBH was first examined by \citet{Peebles72} and then later examined in the context of globular cluster environments \citep{BahcallWolf76, BahcallWolf77}. The authors found that the stellar distribution of stars around a MBH scales as the power-law of the distance from the MBH. As the distance to the MBH approaches zero, the density diverges. \citet{LightmanShapiro77} proposed that MBHs could form in the centres of globular clusters during core collapse from the BH's consumption of stars. \citet{FrankRees76} and \citet{LightmanShapiro77} investigated the rate of tidal disruption events (TDEs) of stars by a MBH in the cores of globular clusters and active galactic nuclei.  

The dynamics of a MBH being perturbed by bound and unbound stars in the core of a globular cluster was first considered by \citet{LinTremaine80}. The authors found that such a MBH undergoes brownian motion due to encounters with unbound stars in the cluster core.  Other authors considered a scenario in which BH binaries at the centres of clusters merge and, due to the anisotropic emission of gravitational waves (GWs), receive a velocity kick that can vary by orders of magnitude \citep{favata04,merritt04,blecha11}. The magnitude of the kick depends on the mass ratio, the magnitude of the BH spins, and the spin inclinations relative to the binary orbital plane.  For very low mass ratios q $\lesssim$ 0.05, the kick velocities are always small $\lesssim$ 100 km/s, independent of the BH spins or their orientations relative to the binary orbital plane (see Figure 1 in \citet{merritt04}). At a mass ratio of q $\sim$ 0.3 for maximally misaligned spins, the kick velocity can be greater than 500 km/s \citep{favata04}.  

Another source of radial kicks is single-binary interactions involving all stellar-mass BHs in the dense cluster core.  Such three-body interactions are chaotic and proceed such that one of the BHs is always ejected, leaving the other two BHs bound as a binary (ignoring dissipative forces; see \citet{valtonen06} and references therein).  The distribution of ejection velocities peaks at several tens of km s$^{-1}$ \citep[e.g.][]{leigh16b}, such that a significant fraction of these ejected BHs will remain bound to their host cluster.

For kicked BHs that remain bound to their host cluster, the trajectory can be described as a damped simple harmonic oscillator \citep{chatterjee02,webb18}. The treatment can be applied to a kicked BH regardless of whether the kick occurs at formation, due to GW emission, or through a single-binary interaction. The decay rate of the BH's orbit does not follow the predictions of classical sDF theory \citep{Chandrasekhar43}, which does not apply to the non-homogeneous stellar field of galaxies and star clusters \citep[e.g.][]{binney77, pesce92, colpi99, vicari07, arca14b}. Newer frameworks have been developed to estimate the behaviour of massive test bodies (i.e BHs) in dense stellar systems, and have successfully been applied to simulations of globular cluster-like environments \citep{antonini12, arca14a, arca16}. Building off these previous studies, \citet{webb18} found that, in the case of a kicked BH, classical sDF theory overestimates the energy loss due to DF in the core of a star cluster. The overestimation stems from the back gravitational focussing of stars by the BH being less effective at the centre of a deep potential (such that stars do not form a wake behind it) compared to a BH passing through a flat distribution of stars.

Furthermore, the Hubble Space Telescope Proper Motion Survey \citep{Bellini17} and the Gaia Satellite \citep{Bianchini18}, show that a large number of star clusters have clear signs of rotation. In fact, over $50\%$ of Galactic GCs have been found to be rotating \citep{Kamann18}. Furthermore, in the 22 GCs that \citet{Bianchini18} detects rotation (out of 51), the authors find that dynamically older clusters rotate slower than dynamically young clusters. The general idea is that present day clusters rotate due to the initial collapse of the giant molecular cloud from which they formed and conservation of angular momentum. Hence, clusters may even have rotated faster in the past, continually losing angular momentum due to internal relaxation \citep{Tiongco17}. Rotation is also not specific to GCs only, as it has been observed in intermediate-age clusters, massive clusters, and nuclear star clusters \citep{HenaultBrunet12,Mackey13, Feldmeier14, Nguyen18}. Therefore, incorporating rotation is crucial to understanding the behaviour of kicked BHs in star clusters, especially for constraining the BH-BH merger rate in star clusters and the production of GW sources.

To date, only the Brownian motion of a BH in a rotating cluster environment has been modeled using a Langevin system of equations by \citet{Lingam18}. However, we expect the orbital decay process to be strongly affected as stars will have a different mean velocity in a preferred direction relative to the BH's motion. The kicked BH will therefore gain angular momentum and start to rotate with the cluster as its orbit decays. The relevant dynamics governing the interplay between these two processes requires further exploration.

This paper, the second in the series after \citet{webb18}, follows the orbital evolution of a MBH in a rotating star cluster environment. Using direct $N$-body simulations, we compare our previous works that studied the linear orbital decay of a MBH in a non-rotating star cluster to our work, where we consider a MBH decaying in a rotating cluster. We then compute the specific energy and angular momentum of the BH as it decays from the linear regime all the way down to the rotating brownian regime, and follow the decay of its amplitude.

\section{N-Body Simulations}

In order to verify our derived relationship between $A_{\rm R}$ and $A_{\rm z}$ and determine how their evolution is related to the properties of both the kicked BH and the rotating cluster, we make use of the direct $N$-body code NBODY6 \citep{aarseth03}. We setup model star clusters with stellar positions and velocities drawn from a Plummer distribution function. Model clusters initially consist of 50,000 star of mass 0.5 $M_\odot$ and have an initial half-mass radius $r_m = 2.5$ pc. In order to consider cluster rotation, we take a fraction q of all stars with $v_{\theta} < 0$ and give them positive rotational velocities about the z-axis \citep{Lynden60}. The stability of rotating clusters setup using this approach has been studied by \citet{Meza02} (see also \citet{Rozier19} for a recent study on the stability of rotating systems). It should be noted that our choice of initial cluster mass and size are such that we can easily compare our simulations to the decay of the kicked BHs in non-rotating clusters presented in \citet{webb18}, where the number of stars was kept purposely low in order to be able to generate a large suite of simulations and perform a systematic study of how kicked BHs behave in star clusters. Once the behaviour of kicked BHs in rotating and non-rotating clusters is understood, we can extend the suite of simulations to include larger cluster masses, a mass spectrum, stellar evolution, and an external tidal field.

Our base set of simulations are setup to maximize the cluster's rotation rate, which optimizes our ability to observe how cluster rotation affects the evolution of kicked BHs since the timescale over which the BH gains angular momentum from cluster stars is minimized. Hence we first consider the case where all stars in the cluster have $v_{\theta} >  0$ (q=1.0). Furthermore, the kicked BH in our base simulations is 100 $M_\odot$ ($0.4\% $of the cluster's total mass) and assumed to be located at the origin of cluster with a kick velocity of 9.6 km/s at time zero. The kick velocity was selected to be approximately twice the value of the cluster's core velocity dispersion $\sigma_c$, ensuring the BH is able to move out beyond the cluster's half-mass radius without escaping. We explore scenarios in which the BH is kicked along the x-axis, in the x-z plane with an inclination of $45^{\circ}$, and along the axis of rotation (z-axis). A non-rotating version of these models was presented in \citet{webb18}, and will be used for comparison purposes in this study. 

\begin{figure}
\begin{center}
\includegraphics[width=\columnwidth]{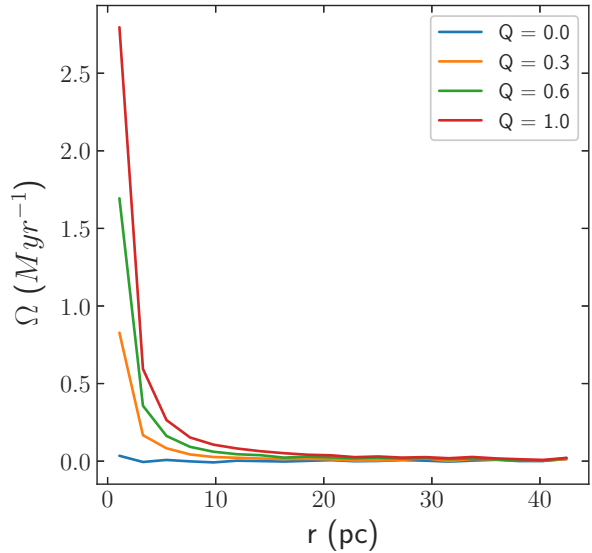}
\end{center}
\caption{The initial mean angular frequency profile of star clusters with q=0.0 (blue), q=0.3 (orange), 0.6 (green), and 1.0 (red) as a function of three dimensional clustercentric distance. \label{fig:oprof}}
\end{figure}

To compare with our initial set of simulations, we also independently vary the BH mass, kick velocity, and degree of rotation. To best compare with the non-rotating cluster models explored in \citet{webb18}, we also consider BH masses of 10 $M_\odot$ and 50 $M_\odot$ and kick velocities ranging from 0.75-1.75 $\sigma_c$. Finally, we consider two intermediate degree's of rotation by setting up model clusters with q=0.3 and q=0.6. It is important to note that clusters with different initial q values are setup using the same set of initial conditions, such that they only differ by the direction of $v_{\theta}$ for select stars. The initial rotation profiles of model clusters produced using this approach are illustrated in Fig \ref{fig:oprof}.

\section{Method} \label{s_method}

In this section, we describe how we compute the specific energy and angular momentum of the kicked BH, and decompose the BH's amplitude in to radial and axial components. To understand the motion of a BH imparted with a kick of magnitude v$_{\rm kick}$ at a clustercentric distance of r $=$ 0 and time t $=$ 0, we assume a Plummer model for the underlying gravitational potential with scale radius a. We further assume the cluster rotates with rotational frequency $\vec{\Omega} = {\Omega}\hat{z}$, which is equal to the mean $\Omega$ of stars within a given BH's maximum kick amplitude. As we will show in Section \ref{trajectory}, a kicked BH primarily gains angular momentum from stars in the core of the cluster (which will dominate the calculation of $\vec{\Omega}$ due to the cluster's density profile). However, since the BH still gains angular momentum over its entire orbit, the mean rotational frequency of stars within the BH's maximum amplitude is a more appropriate choice for $\vec{\Omega}$ over the rotational frequency of core stars. These assumptions allow for a simple analytic expression for the gravitational potential $\Psi$ of a rotating Plummer model to be derived.  

\subsection{Energy and angular momentum of the kicked BH in a rotating cluster} \label{enandang}

Following the derivation in Appendix C of \citet{Lingam18}, we work with a modified distribution function of the form $f(\epsilon - \vec{\Omega} \cdot \ \vec{J})$, where $\epsilon$ is stellar energy and $\vec{J}$ is the angular momentum vector.  The density $\rho$ is then written as:
\begin{equation}
\label{eqn:dens}
\rho = \int f(\epsilon - \vec{\Omega} \cdot \vec{J})d^3v,
\end{equation}
which becomes:
\begin{equation}
\label{eqn:dens2}
\rho = 4\pi \int_0^{\sqrt{2\Psi_{\rm eff}}} f(\Psi_{\rm eff} - \frac{1}{2}v_{\rm 0}'^2)v_{\rm 0}'^2dv_{\rm 0}',
\end{equation}
where the effective potential $\Psi_{\rm eff} = \Psi + \frac{1}{2}v_{\rm rot}^2$ and we have performed a substitution of variables $v_{\rm 0}' = v_{\rm 0} - v_{\rm rot}$. Here $v_{\rm rot}$ is the cluster's rotation velocity. We have also used the fact that $ f = 0$ when its argument is negative, and $\Psi = -\Phi$, where $\Phi$ is the self-consistent gravitational potential in the non-rotating case.  Equation~\ref{eqn:dens2} has the convenient feature that the distribution function is now isotropic in velocity space in the primed coordinates and is identical to the non-rotating case under the substitution $\Psi_{\rm eff} \rightarrow \Psi$.

Expanding $\Psi _{\rm eff}$ yields the rotationally corrected gravitational potential $\Psi + [\Omega^2R^2]/2$ in cylindrical coordinates (R, $\theta$, $\phi$).  The expansion implies that the specific energy of the kicked BH at any location in the potential is:
\begin{equation}
\label{eqn:energy}
E = \frac{1}{2}(v_{\rm R}^2 + v_{\rm \theta}^2 + v_{\rm z}^2) + \Phi_{\rm 0}(R,z) + \frac{1}{2}\Omega^2R^2,
\end{equation}
where $\Phi_{\rm 0}$ is the gravitational potential in the non-rotating case, or:
\begin{equation}
\label{eqn:Phi}
\Phi_{\rm 0} = -\frac{GM}{((R^2+z^2) + a^2)^{1/2}},
\end{equation}
where $M$ is the total cluster mass and $a$ is the Plummer radius.  

Similarly, the specific angular momentum in the z-direction of the kicked BH is:
\begin{equation}
\label{eqn:Lz2}
\vec{L_{\rm z}} = R \vec{v_{\rm \theta}},
\end{equation}

\subsubsection{Amplitude} \label{amp}

The time evolution of the BH's amplitude $A$, which is the maximum orbital distance for a given orbital energy, can be quantified in R and z as follows.  First, invoking conservation of energy, at each time-step we have:
\begin{equation}
\label{eqn:econserve1}
E_{\rm BH} = \frac{1}{2} A_{\rm R} v_{\rm \theta}^2 - \frac{GM}{\sqrt{A_{\rm R}^2 + A_{\rm z}^2 + a^2}} + \frac{1}{2}\Omega^2A_{\rm R}^2,
\end{equation}

where $E_{\rm BH}$ is the total energy of the kicked BH per unit mass, $A_{\rm R}$ is the radial amplitude and $A_{\rm z}$ is the z-component of the amplitude. The 3-D amplitude $A$ is then simply
\begin{equation}
\label{eqn:amp}
A^2 = A_{\rm R}^2 + A_{\rm z}^2,
\end{equation}  

When the BH is at apocentre, it should be noted that $v_{\rm R} = 0$ when $R = A_{\rm R}$ and $v_{\rm z} = 0$ when $z = A_{\rm z}$. Furthermore, Equation~\ref{eqn:Lz2} becomes:

\begin{equation}
\label{eqn:Lz3}
\vec{L_{\rm z}} = A_{\rm R} \vec{v_{\rm \theta}(A)}.
\end{equation}

In order to solve for these parameters at a given moment in time, we can invoke conservation of angular momentum. Hence we can set Equation~\ref{eqn:Lz3} equal to the value of $L_{\rm z,BH}$ at each time-step and solve for $v_{\rm \theta}$ in terms of $A_{\rm R}$. Thus, at every time-step, we have an equation that relates the amplitudes in the R- and z-directions. Once the BH has reached a circular orbit with an orbital frequency equal to that of the cluster, we then simply have two equations (Equations \ref{eqn:econserve1} and \ref{eqn:Lz3}) and two unknowns ($A_{\rm R}$ and $A_{\rm z}$) for which we can solve.

\section{Results} \label{s_results}

In this section, we present the results of $N$-body simulations of kicked BHs in rotating star clusters.  We initially focus on the decay of the BH's oscillatory motion in a rotating cluster potential with q=1 by considering the overall decay rate of its oscillatory amplitude. The analysis is performed on $100 M_{\odot}$ BHs that receive initial kicks of $2 \sigma_c$aligned with the x-axis, the z-axis, and inclined at a 45$^{\circ}$ angle relative to the x- and z-axes. We then separately consider how different rotation rates, BH masses, and kick velocities influence our findings.

\subsection{The trajectory of the BH through its host cluster} \label{trajectory}

\begin{figure}
\begin{center}
\includegraphics[width=\columnwidth]{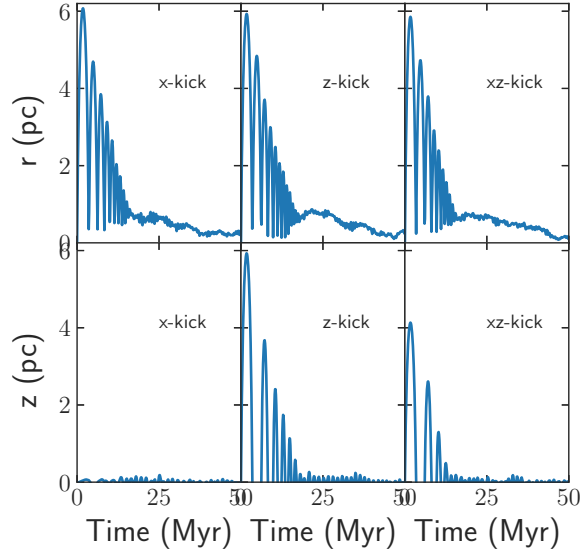}
\end{center}
\caption[Evolution of each kicked BH through its host cluster]{The evolution of each kicked BH through its host cluster as functions of R (top panels) and z (bottom panels) for all three kick scenarios described in Section~\ref{s_method}. In all three cases, the BH's orbit eventually circularizes in the xy-plane due to the cluster rotating about the z-axis.}
\label{fig:xyzplot}
\end{figure}

\begin{figure}
\begin{center}
\includegraphics[width=\columnwidth]{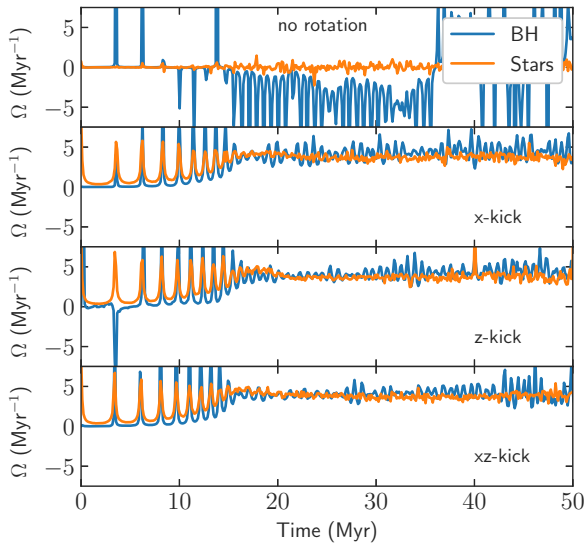}
\end{center}
\caption[The orbital frequency of each kicked BH through its host cluster]{The orbital frequency of each kicked BH (blue) and of nearby stars (orange) over a period $>$ 100 Myr for the non-rotating cluster case (top panel) and all three kick scenarios described in Section~\ref{s_method}. In the rotating cases, the BH gains angular momentum until its orbit circularizes and it reaches the same orbital frequency as local stars.
\label{fig:omega}}
\end{figure}

In Figure~\ref{fig:xyzplot} we show the R and z trajectory of BHs in each of the three kick scenarios described in the previous section. As is clear from Figure~\ref{fig:xyzplot}, the kicked BH's trajectory is non-negligibly affected by the rotation of its host cluster.  Specifically, it ends up on a circular orbit in the xy-plane. To explore the evolution of the BH's orbital frequency, in Figure \ref{fig:omega} we compare the orbital frequency of each BH along its orbit to the time-averaged orbital frequency of stars at the same clustercentric distance. A BH kicked along the x-axis of a non-rotating cluster is shown for comparison purposes. Initially after the BH is first kicked, when it has a near zero orbital frequency, it spends the majority of its time in the outer regions of the cluster where stars also have low orbital frequencies. However as the BH's orbit starts to decay the mean orbital frequency of the background stars will increase, as does that of the BH. It can also be seen by comparing Figures \ref{fig:xyzplot} and \ref{fig:omega} that circularization occurs when the BH's angular frequency is roughly equal to that of the local stars orbiting at nearly the same clustercentric distance. This observation immediately implies that significant angular momentum is imparted to the kicked BH in the rotating cluster case, which is in contrast with what we found for non-rotating spherically-symmetric clusters in \citet{webb18}.  

\subsection{Time evolution of each component of the kicked BH's oscillatory amplitude} \label{ampev}

\begin{figure}
\begin{center}
\includegraphics[width=\columnwidth]{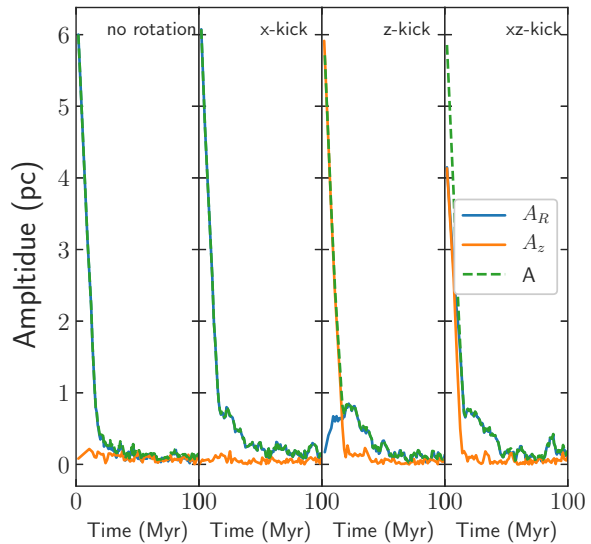}
\end{center}
\caption[The amplitudes for each simulated kicked BH in R- and z-coordinates.]{The amplitude (green dashed lines) over a period of 100 Myr of kicked BHs for the non-rotating case (left panel) as well as all three kick scenarios described in Section~\ref{s_method}. The blue and orange curves show the evolution of the amplitudes in the R- and z-coordinates, respectively. In the rotating cluster cases, the BH's amplitude initially decays due to dynamical friction until its orbit has circularized and the relative velocity between the BH and local stars becomes small.
\label{fig:amp}}
\end{figure}

In Figure~\ref{fig:amp}, we plot the trajectory of the three BH kick scenarios over a period of 100 Myr. The blue and orange curves show the trajectories (and hence evolution of the amplitudes) for the R- and z-coordinates, respectively, and the dotted green curves show the total $r_{\rm BH}^2 = R_{\rm BH}^2 + z_{\rm BH}^2$. In the non-rotating cluster case, the BH has an initial amplitude of $A=6$ pc that then decays linearly with time at a rate of approximately 0.35 pc / Myr. However, as is clear from Figure~\ref{fig:amp}, kicked BHs in rotating clusters consistently gain angular momentum in the z-direction (i.e., $L_{\rm z}$), the cluster's axis of rotation. Interestingly, a new behaviour is apparent in the time evolution of the BH's trajectory through the cluster if it is rotating. In particular, there now appears to be three separate phases in the decay of the kicked BH.  First, the BH follows a roughly linear decay in its amplitude before significantly feeling the effects of the cluster's rotation.  Second, the BH reaches an approximately circular orbit in the xy-plane, with a small amplitude in the z-direction.  The amplitude of this circular orbit then seems to decay approximately linearly in time, but at a much slower rate compared to when its orbit was still eccentric. For the rotating clusters considered here, the average decay rate during this phase is roughly 0.03 pc/Myr, which is an order of magnitude slower than the initial linear orbital decay. Third, once the circular regime has reached sufficiently small distances from the centre of the cluster, the Brownian regime becomes important and rather quickly takes over as describing the time evolution of the BH's motion.  

\begin{figure}
\begin{center}
\includegraphics[width=\columnwidth]{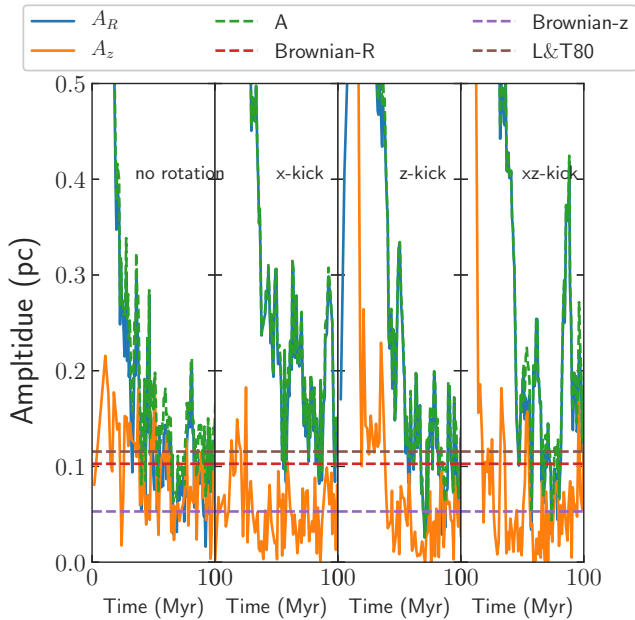}
\end{center}
\caption[The amplitudes for each simulated kicked BH in x-, y- and z-coordinates, zoomed in close to the origin.]{The same as Figure~\ref{fig:amp}, but zoomed in closer to the origin, to highlight the circular and Brownian phases of orbital decay.  The predictions of \citet{Lingam18} for the time-averaged amplitudes of the BH in each spatial direction in the Brownian regime are shown by the horizontal red (R) and purple (z-axis) dashed lines.  The horizontal dashed brown line shows the prediction of \citet{LinTremaine80} for the critical distance from the origin at which the Brownian regime has been reached.  
\label{fig:amp_zoom}}
\end{figure}

In Figure~\ref{fig:amp_zoom} we present a zoomed-in version of Figure \ref{fig:amp} in order to study the later phases of the BH's orbital decay, as the BH transitions from a linear orbital decay to the circular orbit phase and Brownian regime. Focusing on the three separate regimes individually, we find:  (1) the initial linear decay remains similar to what we found in \citet{webb18}, as the amplitude evolution of the BHs in rotating clusters follows that of the non-rotating cluster; (2) the orbital decay is slowed, but remains linear, when the BH reaches a circular orbit (which we call the "circularization radius", or the distance from the cluster centre at which the BH first reaches an approximately circular orbit); that is, when its orbital frequency is the same as the local rotational frequency of stars in the host cluster (see Figure \ref{fig:omega}); and finally (3) the BH's behaviour in the Brownian regime is well described by the previous analytic work of \citet{Lingam18} to within a factor of a few, as shown by the over-plotted red and blue dotted lines in Figure~\ref{fig:xyzplot}. In each case, the R and z decay of the BH's amplitudes appears to stop completely once it reaches the Brownian regime amplitudes predicted by \citet{Lingam18}. Similarily, the total amplitude A agrees quite well to the dashed brown line, which shows the prediction of \citet{LinTremaine80} for the critical distance from the origin at which the Brownian regime has been reached.

Having identified a new phase in the decay of a kicked BH's orbit due to rotation, namely the circularization phase, it is important to next consider if the occurence and duration of this phase is dependent on any cluster or BH properties. We have already shown that the circularization phase occurs regardless of the orientation between the direction of the BH's velocity kick and the rotation axis. To first order, it appears the evolution of the total amplitude $A$ is identical in each case. However to quantify the co-evolution of $A_{\rm R}$ and $A_{\rm z}$, which strongly depend on the kick velocity direction, significantly more intermediate directions would need to be considered. In the following sections we consider how different cluster rotation rates, BH masses, and BH kick velocities affect this scenario in order to identify the relevant parameter space required of large set of models.

\subsection{Time evolution of the kicked BH's oscillatory amplitude for BHs of different mass}

\begin{figure}
\begin{center}
\includegraphics[width=\columnwidth]{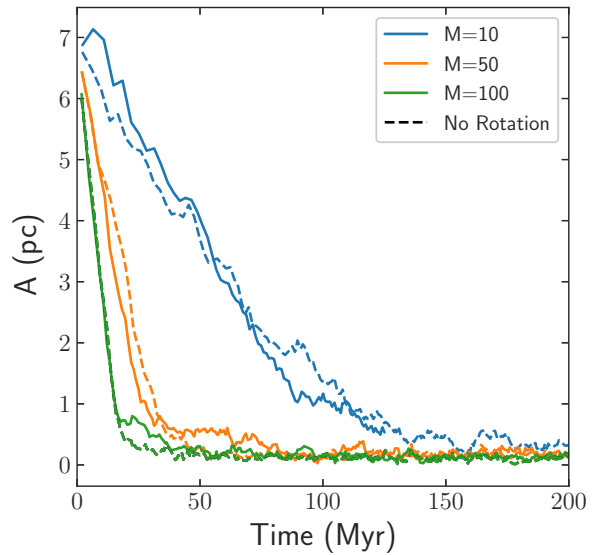}
\end{center}
\caption{The amplitudes for simulated kicked BHs with masses of 10 (blue), 50 (yellow) and 100 $M_\odot$ (green). Solid lines mark clusters rotating with q=1. and dash lines mark non-rotating clusters with q=0.
\label{fig:mplot}}
\end{figure}

Since the strength of the force due to dynamical friction acting on a kicked BH scales linearly with its mass \citep{Chandrasekhar43}, assuming background stars with a fixed mean mass, it is important to consider how the decay of different mass BHs behaves relative to the 100 $M_\odot$ BH presented in Figure \ref{fig:amp}. Therefore we present the results of two additional simulations in Figure \ref{fig:mplot}, where BHs with masses of 10 $M_\odot$ and 50 $M_\odot$ are kicked from the origin of a rotating cluster (q=1.0) with a velocity of 9.6 km/s along the x-axis.

Figure \ref{fig:vplot} illustrates that for a range of BH masses, rotation is still able to circularize the BH's orbit and slow its decay. For the 50 $M_\odot$ and 100 $M_\odot$ BHs, the departure from the non-rotating cases is evident, with the decay of the BH's amplitude stalling for several tens of Myr before slowly decaying into the cluster's core. For the 10 $M_\odot$ BH, while the BH's orbit circularizes around 90 Myr it still reaches the core of the cluster at the same time as the non-rotating case. The lack of a discrepancy is due to the 10 $M_\odot$ BH in the non-rotating simulation undergoing a close encounter at $\sim$ 90 Myr which results in an energy gain (as discussed in \citet{webb18}). Lower mass BH's are more strongly affected by close encounters than high mass BHs. Hence lower mass BHs can have similar overall decay times in rotating and non-rotating clusters due to close encounters, however only in the rotating case is the longer decay time due to a circularization of the BHs orbit followed by a slow decay into the core.

\subsection{Time evolution of the kicked BH's oscillatory amplitude for different initial kick velocities}

\begin{figure*}
\begin{center}
\includegraphics[width=\textwidth]{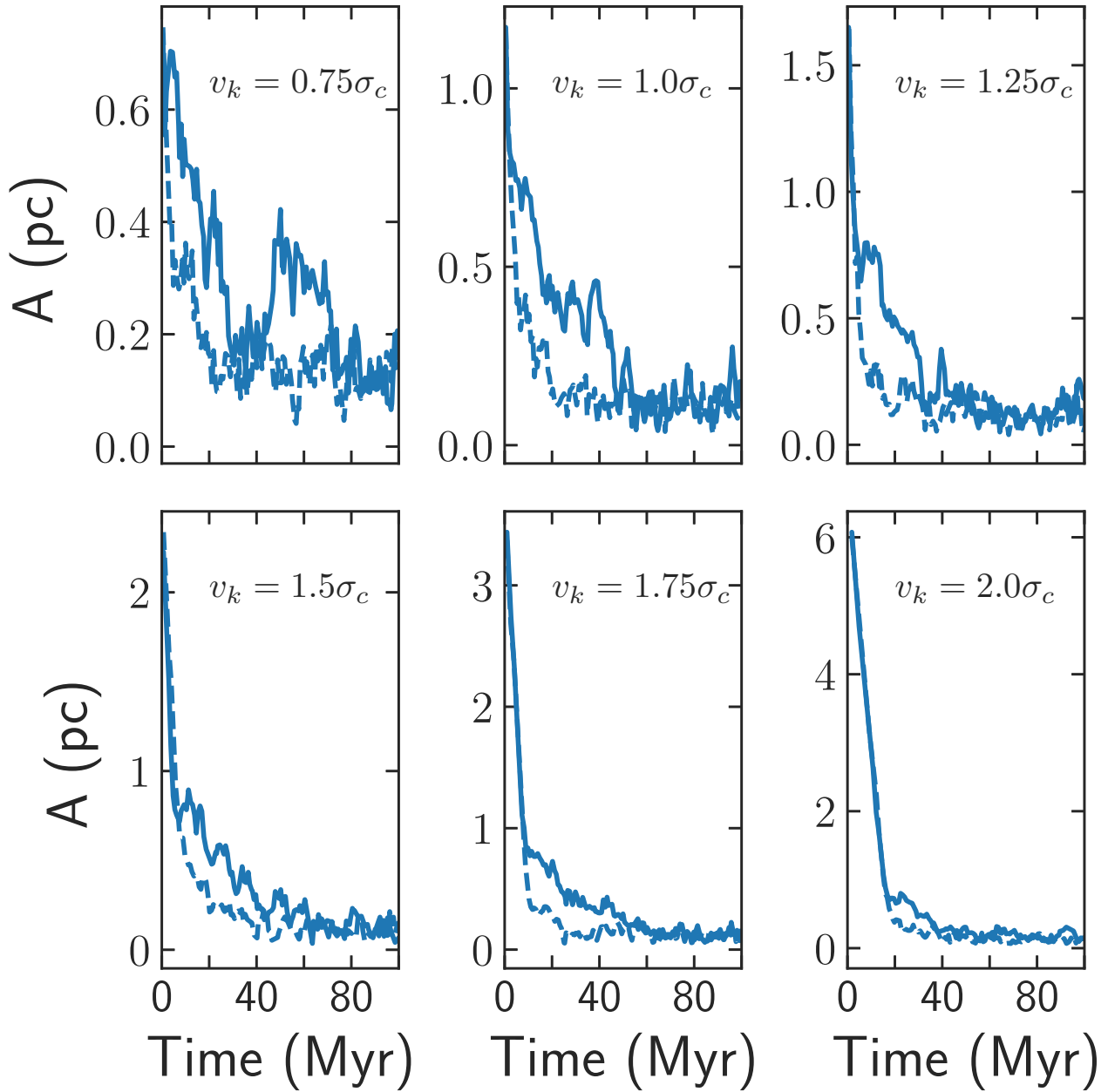}
\end{center}
\caption{The amplitudes for simulated kicked BHs with kick velocities ranging from 0.75 to 2 $\times$ $\sigma_c$. Solid lines mark clusters rotating with q=1. and dash lines mark non-rotating clusters with q=0.
\label{fig:vplot}}
\end{figure*}

A kicked BHs orbital decay time will also depend on the kick velocity itself, as BHs that are kicked farther from the centre of the cluster will take longer to decay back to the centre for two reasons. First, since the stellar density is lower in the outer regions of a cluster the effects of dynamical friction will be weaker. In fact, in \citet{webb18} we partially attribute classical dynamical friction`s inability to predict the decay of BHs that recieve large velocity kicks to the the varying background stellar density they experience. Second, the BH will have a higher initial kinetic energy such that it must lose more energy before being able to reach the Brownian motion phase inside the cluster's core. To explore how rotation effects these two processes, we re-simulate the model clusters with 100 $M_\odot$ BHs kicked along the x-axis with kick velocities between 0.75 and 1.75 $\times$ $\sigma_c$. The orbital decay of the BHs in rotating clusters are compared directly to BHs in non-rotating clusters from \citet{webb18} in Figure \ref{fig:vplot}.

\begin{figure*}
\begin{center}
\includegraphics[width=\textwidth]{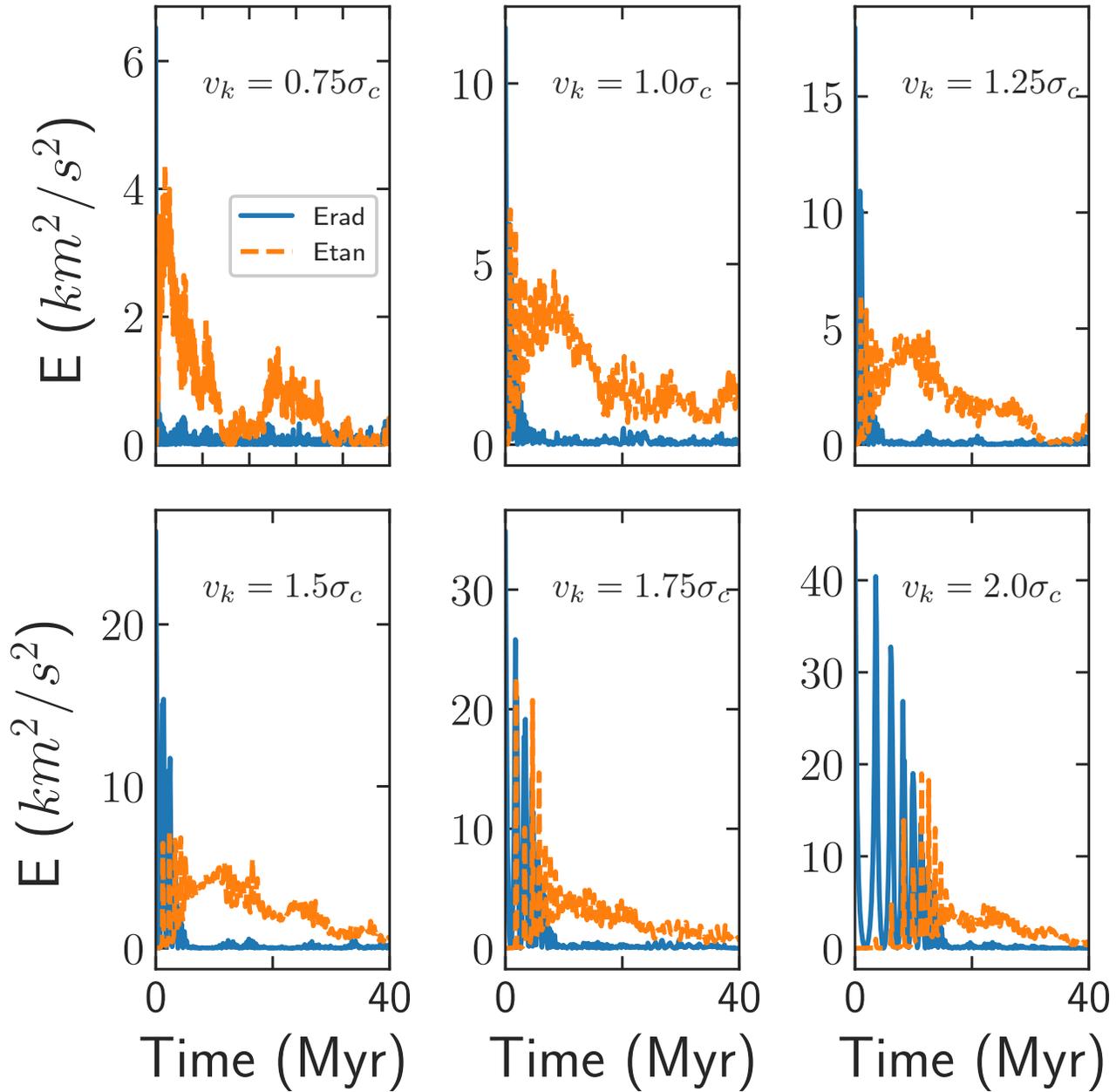}
\end{center}
\caption{The radial and tangential kinetic energies for simulated kicked BHs with kick velocities ranging from 0.75 to 2 $\times$ $\sigma_c$. Solid lines mark clusters rotating with q=1. and dash lines mark non-rotating clusters with q=0.
\label{fig:ek_vplot}}
\end{figure*}

For low kick velocities, Figure \ref{fig:vplot} illustrates that kicked BHs will decay quickly to the clusters centre and do not have time to gain angular momentum from rotating stars. Hence their orbital decay is similar to the non-rotating case described in \citet{webb18}. However for intermediate kick velocities (1-1.75 $\sigma_c$), kicked BHs are in fact more sensitive to the fact that the host cluster is rotating. As illustrated in Figure \ref{fig:ek_vplot}, a lower initial radial amplitude means it takes less time for tangential motion gained by interacting with rotating stars to overcome a BH's radial motion. Hence, since lower kick velocities result in BHs remaining in the dense inner regions of a GC where the rotation rate is high, the circularization timescale will both be shorter. Quickly reaching the circular orbit phase causes BHs that receive kicks of 1.5 and 1.75 $\sigma_c$ to take even longer to decay in a rotating cluster than a BH that receives a 2 $\sigma_c$, despite their lower initial amplitudes.

\subsection{Time evolution of the kicked BH's oscillatory amplitude for different cluster rotation rates}

\begin{figure}
\begin{center}
\includegraphics[width=\columnwidth]{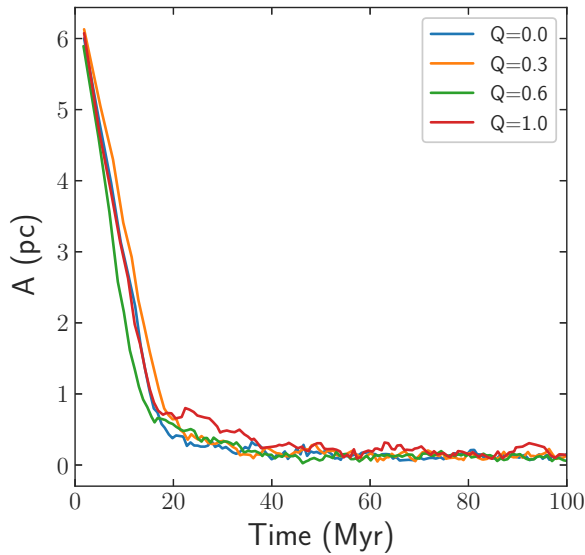}
\end{center}
\caption{The amplitudes for simulated kicked BHs in clusters rotating with q=0 (blue), 0.3 (orange), 0.6 (green) and 1.0 (red).
\label{fig:qplot}}
\end{figure}

As previously discussed, we initially set our cluster's to be rotating with q=1 to maximize the effect of rotation on the orbital decay of kicked BHs. Figure \ref{fig:qplot} illustrates the decay of BHs kicked with initial velocities of $2 \sigma_c$ along the x-axis of clusters with q=0, 0.3, 0.6, and 1 about the z-axis. For all four cases the initial linear decay phases are comparable, with some minor differences due to the fact that changing the sign of $v_{\theta}$ for select stars results in the clusters quickly becoming different realizations of each other. At later times, the low rotation rate of the q=0.3 cluster appears to have a negligible effect on the BH's decay. Hence, for the relatively high-mass BHs considered here, there appears to be a minimum rotation rate that a cluster must have before the orbital decay of BHs are affected.

The BH in the q=0.6 cluster on the other hand appears to reach circularization at a slightly lesser radius than the q=1 models initially presented. The decay of the BH after circularization is also faster in the q=0.6 than the q=1., indicating rotation plays a role in the post-circularization decay as well. Since counter-rotating stars that interact with the BH will have high relative velocities, they do not contribute to the dynamical friction force and decrease the rate at which the BH gains angular momentum \cite{gualandris08}.

{}

While the decay of the BH in the q=0.3 model cluster is unaffected by the fact that the cluster is rotating, it is still gaining angular momentum as its orbit decays. There is simply an interplay between rotation rate, cluster density, BH mass, and initial kick velocity that determines if the decay will slow before the BH reaches the distance at which Brownian motion governs the BHs trajectory. For example, extending our suite of simulations to higher mass and denser clusters will likely increase how strongly a kicked BH is affected by even low rotation rates as the effects of dynamical friction will be maximized. Such clusters will also allow for larger kick velocities and higher initial amplitudes, giving a kicked BH more time to reach circularization before it reaches the cluster's core. BHs with different masses will also decay differently in rotating clusters, with more massive BHs being more strongly affected by dynamical friction.

\section{Discussion}\label{s_discussion}

Our simulations have revealed that the orbits of kicked BHs in rotating star clusters decay differently than if the host cluster was not rotating. The key difference being that in a rotating cluster, a BHs orbit can circularize if it is able to gain enough angular momentum before reaching the Brownian motion regime. Once circularization occurs, the decay of the BH's orbit due to dynamical friction slows significantly. In the following sections we will discuss the circularization process in more detail, explore how the BH's decay after circularization differs from before circularization, and predict the ramifications that circularization has on the cluster itself.


\subsection{Circularization}

\begin{figure}
\begin{center}
\includegraphics[width=\columnwidth]{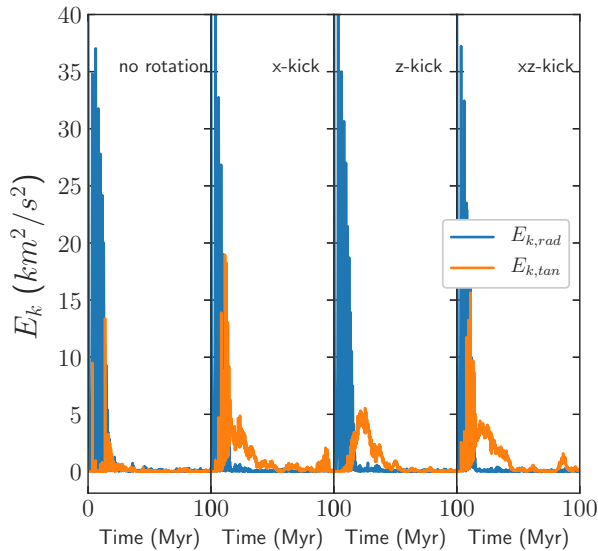}
\end{center}
\caption[The radial (blue) and tangential (orange) kinetic energies for each simulated kicked BH.]{The radial (blue) and tangential (orange) kinetic energies over a period of 100 Myr for kicked BHs in the non-rotating case (left panel) as well as all three kick scenarios described in Section~\ref{s_method}, namely initial kicks aligned with the x-axis (left center panel), the z-axis (right center panel) and initial kicks inclined at a 45$^{\circ}$ angle relative to the x- and z-axes (right panel). 
\label{fig:ekrat}}
\end{figure}

The circularization process is best illustrated by considering the radial and tangential kinetic energy evolution of BHs received $2 \sigma_c$ velocity kicks in different directions in rotating model clusters with q=1. As illustrated in Figure \ref{fig:ekrat}, once the BH gains enough angular momentum from its host cluster, such that its tangential kinetic energy becomes greater than its radial kinetic energy, its orbit will circularize and it will have the same orbital frequency as local stars. A near-zero relative velocity between the BH and local stars renders the effects of dynamical friction negligible \citep{Chandrasekhar43}, which explains the change in the BH's decay rate. It should be noted, however, that these simulations have been setup to minimize the timescale over which the BH's orbit will circularize and maximize the BH's orbital decay time. Hence for different combinations of the BH's mass (see Figure \ref{fig:mplot}), the initial kick velocity (see Figure \ref{fig:vplot}), and the cluster's rotation rate (see Figure \ref{fig:qplot}, it is possible that the BH's orbit can fully decay before circularizing or circularize even earlier than the models in Figure \ref{fig:ekrat}.

To estimate the fraction of the cluster's angular momentum taken by the BH, we integrate numerically over our rotating Plummer potential out to the initial maximum amplitude. For cluster's rotating with q=1, we find the total initial angular momentum of stars within the maximum amplitude to be between 
2.6 $\times$ 10$^3$ and 5.6 $\times$ 10$^5$ M$_{\odot}$ $pc^2$ Myr$^{-1}$ for kick velocities between 0.75 and 2 $\sigma_c$. At the circularization radius, model BHs with different kick velocities have total angular momenta between 1.5 and 2.5 $\times$ 10$^2$ M$_{\odot}$ $pc^2$ Myr$^{-1}$. Thus, by the time its amplitude has decayed to the circularization radius, the BH will have removed between 0.04 and 0.6$\%$ of the total angular momentum within the maximum kick amplitude for the range of initial conditions considered here. BHs that receive lower kick velocities remove a higher fraction of the cluster's angular momentum as they spend a longer time in the inner region where $\Omega$ is highest. The total angular momentum of stars within the maximum amplitude will scale linearly with q and the fraction of angular momentum removed from the cluster by the BH scales inversely with q. In order to decay to the origin, the BH must then re-distribute this angular momentum back to the cluster via two-body relaxation within the circularization radius. Hence, the BH can be kicked out of the core of order 10$^3$ times before significantly affecting the cluster's structure. A population of BHs on the other-hand will of course have a stronger effect on cluster structure, as each BH will re-distribute its angular momentum back into the inner regions of the cluster. However, to quantify the effect, assumptions need to be made regarding the distribution of BH kick velocities and the escape rate of BHs due to two-body and three-body interactions.

It is important to note that low cluster rotation rates or low initial kick velocities can result in BHs decaying to the Brownian regime before circularization. However for intermediate kick velocities and lower BH masses, the BH reaches circularization and its orbital decay slows in a qualitatively similar way as discussed above. Taking into consideration the full suite of simulations, which range in cluster rotation rate, BH mass, kick velocity, and kick direction, a BH can remove between 0.02 and 0.07$\%$ of the cluster's angular momentum during its first decay. Extending the suite of simulations to GC-like masses and densities may result in circularization occurring even quicker as the dynamical friction force is stronger in denser environments.

\subsection{Orbital Decay After Circularization}

\begin{figure}
\begin{center}
\includegraphics[width=\columnwidth]{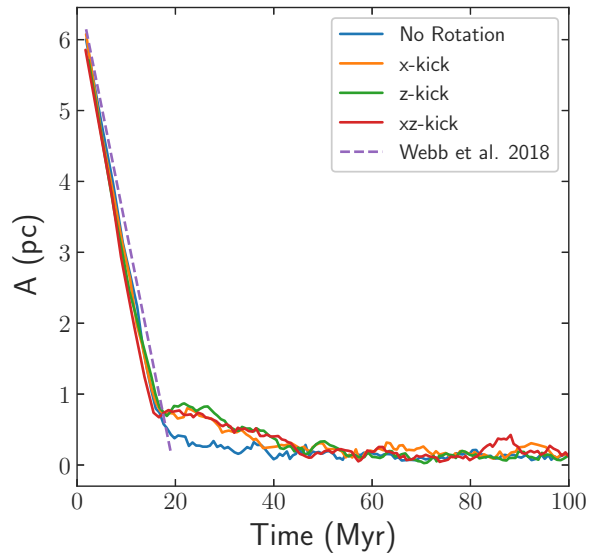}
\end{center}
\caption[The amplitudes for each simulated kicked BH in R- and z-coordinates.]{The amplitude over a period of 100 Myr of kicked BHs for the non-rotating case (black) and all three kick scenarios described in Section~\ref{s_method}, namely initial kicks aligned with the x-axis (bluel), the z-axis (red) and initial kicks inclined at a 45$^{\circ}$ angle relative to the x- and z-axes (yellow).  The dotted black line shows the fit to the non-rotating case from \citet{webb18}, which also fits the kicked BHs in rotating clusters until their orbits circularize at the same orbital frequency as the cluster.
\label{fig:fit}}
\end{figure}

Once circularization happens, the BH will have a velocity that is comparable to nearby stars. In this scenario the effects of dynamical friction are negligible \citep{Weinberg86}, which is why the linear decay formalism from \citet{webb18} (which already predicts longer dynamical friction times than \citet{Chandrasekhar43}) fails to reproduce the BH's decay after circularization. Figure \ref{fig:fit} shows fits using \citet{webb18} to the evolution of $100 M_{\odot}$ BHs that receive $2 \sigma_c$ velocity kicks in model clusters with q=1). In fact, the BH decays at a rate that is approximately an order of magnitude slower after circularization than before circularization. A slower post-circularization decay is also seen when using lower mass BHs or intermediate kick velocities, with the latter case yielding the largest difference between the decay times of BHs in rotating and non-rotating clusters. Hence an entirely different mechanism must be responsible for the kicked BH's evolution as its orbit continues to decay into the Brownian regime.

After the BH's orbit has circularized, the dominant mechanism behind its evolution becomes two-body interactions. Hence the BH's orbit will only decay due to it transferring kinetic energy to low-mass stars as the cluster's core evolves towards a state of partial energy equipartition \citep[e.g][]{trenti13,bianchini16,spera16}. For the models considered in Figure \ref{fig:fit}, circularization occurs once the BH's orbital amplitude reaches approximately 0.75 pc. At this clustercentric distance, the local relaxation time for an object $200 \times$ more massive than the mean mass of local stars would be approximately 2 Myr \citep{binney87} if the cluster was not rotating. Therefore we attribute the fact that it takes nearly 30 Myr for the BH's orbit to decay from 0.75 pc to 0.2 pc (when Brownian motion dominates the BH's evolution) to two-body interactions in a rotating cluster. As noted by \citet{LongarettiLagoute96}, rotation can significantly increase a cluster's relaxation time (see also \citet{Ernst07}). In fact, \citet{LongarettiLagoute96} find that in cases where the ratio of the cluster's total kinetic energy to its rotational energy is $\sim 0.25$, the actual relaxation time can be up to a factor of $3 \times$ longer than in the non-rotating case. In the q=1 models considered here the ratio of the cluster's total kinetic energy to its rotational energy is closer to 0.5, much higher than the range considered in \citet{LongarettiLagoute96}. 

Finally, we note that, since the BH and local stars all have comparable orbital frequencies at the circularization radius, their orbits are in resonance and this affects the decay rate \citep{tremaine84}. Hence the BH's decay rate will also be affected by correlated encounters rather than just the usual two-body relaxation rate for which the encounters are uncorrelated. As we discuss in the next Section, understanding how the BH's decay time is delayed will have an important impact on a kicked BH's ability to form a binary.

\subsection{Tidal Capture}
The fact that BHs in rotating clusters take longer to decay increases the chances that the BH will capture a stellar companion on its way to the origin. This is because, as shown in Figure~\ref{fig:omega}, the relative velocity between the BH and nearby stars is very low at the circularization radius (which we take to be $\sim$ 0.75 pc; see Figures~\ref{fig:xyzplot} and~\ref{fig:amp}). To calculate the capture time-scale we use Equation 5 in \citet{kalogera04} and assume a mean stellar number density of $n = 1.8 \times 10^3$ pc$^{-3}$ (obtained from our simulations at the local circularization radius of the BH, defined as the distance from the cluster centre at which the BH first reaches an approximately circular orbit in its host cluster; see Figures~\ref{fig:xyzplot} and~\ref{fig:amp}), a relative velocity between the BH and nearby stars of $v_{\rm rel} = 1.5$ km s$^{-1}$ (computed from Figure~\ref{fig:omega}; note that we replace the velocity dispersion in Equation 5 of \citet{kalogera04} with the relative velocity), a BH mass of 100 M$_{\rm \odot}$ and a tidal capture radius of $R_{\rm TC} = 10$ R$_{\rm \odot}$. Combining all these variables we estimate that the predicted time-scale for tidal capture of stars to occur by the BH, once it has reached the circular regime, to be on the order of 400 Myr. With our kicked BHs taking approximately 40 Myr to decay, over half of which occurs in the cluster's core after the BH's orbit has circularized, the chance of tidal capture is then approximately $10\%$ over the course of the BH's decay in to the Brownian regime. There is approximately no chance of tidal capture occurring in the linear non-rotating cluster case, since the BH always passes through the core with high relative velocity, at least until it reaches the Brownian regime, and the decay time is much faster.

The take-away message from this simple calculation is that, since the BH ends up on a roughly circular orbit with approximately the same orbital frequency as the other stars in its immediate vicinity, then the relative velocities between the stars and the BH are typically very low.  \textit{The low relative velocity decreases the tidal capture and/or disruption time-scale, such that it is comparable to the duration of the circular phase of the BH's orbital decay.} The chances of the BH directly capturing another BH via GW emission during the circularization is however rather low, as it would require two BHs of comparable mass to receive identical velocity kicks at the exact same time. BH-BH binaries can still form, however, via subsequent exchange encounters with any MS-BH binaries that form due to tidal capture.  The formation of BH-BH binaries is therefore relegated to the Brownian phase when a BH-main sequence (MS) star binary will undergo exchanges with other BHs and BH-MS star binaries. 

With the above said, we strongly caution that there are many unknowns in performing more detailed calculations of this scenario.  In particular, it is unlikely that the BHs can be much more massive than the MS stars they tidally capture, but the exact BH mass above which tidal capture can no longer occur, and tidal disruption will always occur instead, is highly uncertain \citep[e.g.][]{generozov18}.  Nevertheless, we emphasize that more typical BH masses closer to 10 M$_{\odot}$ \citep{spera15,spera17} could actually correspond to a higher probability of tidal capture in the circular decay regime discussed above, since the lower BH mass should correspond to a longer damping time and hence a prolonged circular decay phase.

Finally, we might naively expect the efficiency of tidal capture to be even higher in denser cluster environments, provided the dispersion about the mean orbital frequency of stars remains small compared to the orbital velocity at the circularization radius.  Globular clusters and nuclear star clusters, for example, have number densities that can be several orders of magnitude higher than our model clusters. Hence kicked BHs in realistic cluster environments may have even shorter tidal capture timescales. Therefore, the predicted number of gravitational wave emission events due to BH-BH mergers in star clusters will be possibly higher than if clusters are assumed to not be rotating.

The simple arguments presented in this section motivate the need for more detailed simulations specifically looking at tidal capture in our model clusters. Simulations of rotating and non-rotating clusters with an initial mass spectrum that leads to a sub-population of retained BHs that can interact with main sequence stars will be particularly useful. From these models, a more quantitative analysis on the effects of rotation on kicked BHs can be performed.


\section{Conclusion}\label{s_conclusion}

In this paper, we consider the time evolution of the trajectory of a stellar-mass BH imparted with a velocity kick at the origin of a rotating Plummer potential.  This is done using $N$-body simulations carried out using the \texttt{NBODY6} code.  In our simulations, the cluster is rotating with an angular frequency $\vec{\Omega} = \Omega\hat{z}$, and the BH is given a range of initial kick velocities that are smaller than the local escape speed.  Such kicks should occur naturally due to natal kicks,  kicks imparted via the anisotropic emission of GWs during the final inspiral of two stellar-mass BHs \citep{favata04,merritt04,blecha11} (which are thought to occur commonly at the centres of dense star clusters \citep[e.g.][]{leigh13a, leigh13b, rodriguez15, rodriguez16, leigh16}), and three-body interactions between BHs in a cluster's core \citep[e.g.][]{valtonen06}. In our simulations, we consider three different initial kick directions, namely along the x-axis, along the z-axis and at 45$^{\circ}$ to both the x- and z-axes and three different cluster rotation rates.  

We follow the time evolution of the kicked BHs in our simulations, in particular their amplitude in both the radial (i.e., $A_{\rm R}$) and axial (i.e., $A_{\rm z}$) directions, due to dynamical friction.  We compute the specific energy and angular momentum of the BH, to quantify the amount of angular momentum transferred from the cluster to the BH before it reaches the cluster's core.  This framework extends our previous treatment of the motion of the BH as a damped 1-D simple harmonic motion in a non-rotating cluster in \citet{webb18} to 2 dimensions to account for cluster rotation. 

The results of these simulations show that for BHs that are kicked outside of the cluster's core, the time evolution of the kicked BH is initially well-described by that of a kicked BH in a non-rotating Plummer potential \citep{webb18}.  As the BH's orbit decays it will also gain angular momentum until it either reaches the Brownian motion phase or develops a roughly circular orbit with approximately the local angular frequency of the cluster. This circularization represents a new phase in the decay of a kicked BH's orbit in dense star cluster environments, relative to the non-rotating case. If the velocity kick and the cluster's rotation rate are high enough such that circularization occurs before the BH reaches the core, the decay of the BH back to the origin will be delayed (relative to the non-rotating case). We further find that BHs that receive intermediate kick velocities are most affected by the cluster's rotation, likely due to the fact that they spend more time in the cluster core where they gain angular momentum from stars with higher relative velocities. It therefore takes less time for a BH that receives an intermediate velocity kick to lose radial kinetic energy and have its orbit circularize than a BH that receives a kick that is near the escape velocity of the cluster. This new phase in the decay of the BH takes of order $\sim 100$ Myr for our initial conditions and assumptions, which is comparable to the BH's relaxation time in a rotating environment \citep[][e.g]{LongarettiLagoute96}. Hence the BH's orbit will decay after circularization due to two-body interactions between the BH and local stars until eventually the BH enters the Brownian regime \citep[e.g.][]{chatterjee02,Lingam18}. 

Our results also suggest that a kicked BH in a rotating cluster environment can have up to an order of magnitude higher probability of tidally capturing (or disrupting) a MS star before decaying back to the origin due to dynamical friction than in a non-rotating cluster, with a probability of about 10\% over the entire duration of the decay.  \textit{Hence lower mass\footnote{Here, we remind the reader that, above a critical but theoretically unknown BH mass, tidal capture is no longer possible and only tidal disruption can occur \citep[e.g.][]{generozov18}} BHs sitting at the centres of rotating star cluster potentials have a significant probability of harbouring MS binary companions.} Observationaly, this finding suggests that rotating star clusters may host a larger number of X-ray binaries than non-rotating clusters of comparable size and mass. Furthermore, rotating clusters will also likely host more BH-BH binaries (and be more likely to produce gravitational wave emission events), as they form via three-body encounters where the MS star in a BH-MS star binary is exchanged for a nearby BH.

The simulations presented in this paper are meant as a benchmark that can be extended to, for example, a wide range of rotating or axisymmetric potentials, host star cluster density profiles, BH masses and BH kick velocities. The relative timescales for BH orbit circularization and BH orbit decay will determine how important the role of rotation is when modelling the evolution of BHs in cluster environments. Including an entire sub-population of BHs in future studies will also be important in order to study BH-BH interactions and the effect that BHs will have on their host rotating cluster. Understanding these timescales both in present-day and high-redshift astrophysical environments is essential in constraining GW emission rates and the formation of MBHs.

\section*{Acknowledgements}
JW acknowledges financial support through a Natural Sciences and Engineering Research Council of Canada (NSERC) Postdoctoral Fellowship and additional financial support from NSERC (funding reference number RGPIN-2015-05235) and an Ontario Early Researcher Award (ER16-12-061). MS acknowledges funding from the European Union’s Horizon 2020 research and innovation program under the Marie-Sklodowska-Curie grant agreement No. 794393. AAT acknowledges support from JSPS KAKENHI Grant Number 17F17764. We would also like to kindly thank Enrico Vesperini for useful discussions and suggestions. This work was made possible in part by the facilities of the Shared Hierarchical Academic Research Computing Network (SHARCNET:www.sharcnet.ca) and Compute/Calcul Canada.

\bsp

\label{lastpage}


\begin{thebibliography}{99}

\bibitem[\protect\citeauthoryear{Aarseth}{2003}]{aarseth03} Aarseth, S.J. 2003, Gravitational $N$-body Simulations: Tools and Algorithms (Cambridge Monographs on Mathematical Physics). Cambridge University Press, Cambridge
\bibitem[{{Alessandrini} {et~al.}(2016)}]{Alessandrini16} Alessandrini, E., Lanzoni, B., Ferraro, F.R., Miocchi, P., Vesperini, E. 2016, ApJ, 833, 252
\bibitem[\protect\citeauthoryear{Antonini et al.}{2012}]{antonini12} Antonini F., Capuzzo-Dolcetta R., Mastrobuono-Battisti A., Merritt D. 2012, ApJ, 750, 111
\bibitem[\protect\citeauthoryear{Arca-Sedda \& Capuzzo-Dolcetta}{2014a}]{arca14a} Arca-Sedda, M. \& Capuzzo-Dolcetta, R. 2014a, ApJ, 785, 51
\bibitem[\protect\citeauthoryear{Arca-Sedda \& Capuzzo-Dolcetta}{2014b}]{arca14b} Arca-Sedda, M. \& Capuzzo-Dolcetta, R. 2014b, MNRAS, 444, 3738
\bibitem[\protect\citeauthoryear{Arca-Sedda}{2016}]{arca16} Arca-Sedda, M. 2016, MNRAS, 455, 25
\bibitem[\protect\citeauthoryear{Bahcall \& Wolf}{1976}]{BahcallWolf76} Bahcall J. N., Wolf R.A., 1976, ApJ, 209, 214
\bibitem[\protect\citeauthoryear{Bahcall \& Wolf}{1977}]{BahcallWolf77} Bahcall J. N., Wolf R.A., 1977, ApJ, 216, 883 
\bibitem[\protect\citeauthoryear{Bellini et al.}{2017}]{Bellini17} Bellini, A., Bianchini, P., Varri, A. L., Anderson, J., Piotto, G., van der Marel, R. P., Vesperini, E., Watkins, L. L. 2017, ApJ, 844, 167
\bibitem[{{Bianchini} {et~al.}(2016)}]{bianchini16} Bianchini, P., van de Ven, G., Norris, M.A., Schinnerer, E. Varri, A.L. 2016, MNRAS, 458, 3644
\bibitem[\protect\citeauthoryear{Bianchini et al.}{2018}]{Bianchini18} Bianchini, P., van der Marel, R. P., del Pino, A., Watkins, L. L., Bellini, A., Fardal, M. A., Libralato, M., Sills, A. 2018, MNRAS, Submitted, arXiv:1806.02580
\bibitem[\protect\citeauthoryear{Binney}{1977}]{binney77} Binney J. 1977, MNRAS, 181, 735
\bibitem[\protect\citeauthoryear{Binney \& Tremaine}{1987}]{binney87}
  Binney J., Tremaine S., 1987, Galactic Dynamics (Princeton:
  Princeton University Press)
\bibitem[\protect\citeauthoryear{Blecha et al.}{2011}]{blecha11} Blecha L., Cox T. J., Loeb A., Hernquist L. 2011, MNRAS, 412, 2154
\bibitem[\protect\citeauthoryear{Chandrasekhar}{1943}]{Chandrasekhar43} Chandrasekar S. 1943, ApJ, 97, 255 
\bibitem[\protect\citeauthoryear{Chatterjee}{2002}]{chatterjee02} Chatterjee P., Hernquist L., Loeb, A. 2002, ApJ, 572, 371
\bibitem[\protect\citeauthoryear{Colpi et al.}{1999}]{colpi99} Colpi, M., Mayer, L., Governato, F. 1999, ApJ, 525, 720
\bibitem[\protect\citeauthoryear{Ernst et al.}{2001}]{Ernst07} Ernst, A., Glaschke, P., Fiestas, J., Just, A., Spurzem, R. 2007, MNRAS, 377, 465



\bibitem[\protect\citeauthoryear{Favata, Hughes \& Holz}{2004}]{favata04} Favata M., Hughes S. A., Holz D. E. 2004, ApJ, 607, L5
\bibitem[\protect\citeauthoryear{Feldmeier et al.}{2014}]{Feldmeier14} Feldmeier A., et al., 2014, A\&A, 570, A2

\bibitem[\protect\citeauthoryear{Frank \& Rees}{1976}]{FrankRees76} Frank F., Rees F. M., 1976, MNRAS, 176, 633
\bibitem[\protect\citeauthoryear{Generozov et al.}{2018}]{generozov18} Generozov A., Stone N. C., Metzger B. D., Ostriker J. P. 2018, MNRAS, 478, 4030 
\bibitem[\protect\citeauthoryear{Gualandris \& Merritt}{2008}]{gualandris08} Gualandris A., Merritt D. 2008, ApJ, 678, 780

\bibitem[\protect\citeauthoryear{H\'{e}nault-Brunet et al.}{2012}]{HenaultBrunet12} H\'{e}nault-Brunet V., et al., 2012, A\&A, 546, A73

\bibitem[\protect\citeauthoryear{Kalogera et al.}{2004}]{kalogera04} Kalogera V., King A. R., Rasio F. A. 2004, ApJL, 601, L171
\bibitem[\protect\citeauthoryear{Kamann et al.}{2018}]{Kamann18} Kamann S., et al., 2018, MNRAS, 473, 5591

\bibitem[\protect\citeauthoryear{Leigh et al.}{2013a}]{leigh13a} Leigh N. W.,
Mastrobuono-Battisti A., Perets H. B., B\"oker T. 2013, MNRAS, 441, 919 (Leigh et al. 2013a)
\bibitem[\protect\citeauthoryear{Leigh et al.}{2013b}]{leigh13b} Leigh N. W. C.,
B\"oker T., Maccarone T. J., Perets H. B. 2013, MNRAS, 429, 2997 (Leigh et al. 2013b)
\bibitem[\protect\citeauthoryear{Leigh, Geller \& Toonen}{2016}]{leigh16} Leigh N. W. C.,Geller A. M., Toonen S. 2016, ApJ, 818, 21
\bibitem[\protect\citeauthoryear{Leigh et al.}{2016}]{leigh16b} Leigh N. W. C., Stone N. C., Geller A. M., Shara M. M., Muddu H., Solano-Oropeza D., Thomas Y. 2016, MNRAS, 463, 3311
\bibitem[\protect\citeauthoryear{Lightman \& Shapiro}{1977}]{LightmanShapiro77} Lightman A. P., Shapiro S. L., 1977, ApJ, 211, 244 
\bibitem[\protect\citeauthoryear{Lin \& Tremaine}{1980}]{LinTremaine80} Lin D. N. C., Tremaine S., 1980, ApJ, 242, 789
\bibitem[\protect\citeauthoryear{Lingam}{2018}]{Lingam18} Lingam M., 2018, MNRAS, 473, 1719
\bibitem[\protect\citeauthoryear{Longaretti \& Lagoute}{1996}]{LongarettiLagoute96} Longaretti, P.-Y. \& Lagoute, C. 1996, A\&A, 308, 453
\bibitem[{{Lutzgendorf} {et~al.}(2013)}]{Lutzgendorf13} Lutzgendorf, N., Baumgardt, H., Kruijssen, J.M.D. 2013, A\&A, 558, A117
\bibitem[\protect\citeauthoryear{Lynden-Bell}{1960}]{Lynden60} Lynden-Bell, D. 1960, MNRAS, 120, 204

\bibitem[\protect\citeauthoryear{Mackey et al.}{2013}]{Mackey13} Mackey A. D., Da Costa G. S., Ferguson A. M. N., Yong D., 2013, ApJ, 762, 65

\bibitem[\protect\citeauthoryear{Merritt}{2004}]{merritt04} Merritt D., Milosavljevic M., Favata M., Hughes S. A., Holz D. E. 2004, ApJL, 607, L9
\bibitem[\protect\citeauthoryear{Meza}{2002}]{Meza02} Mezza A. 2002, A\&A, 395, 25M
\bibitem[\protect\citeauthoryear{Nguyen et al.}{2018}]{Nguyen18} Nguyen D. D., et al., 2018, ApJ, 858, 118

\bibitem[\protect\citeauthoryear{Peebles}{1972}]{Peebles72} Peebles P. J. E., 1972, ApJ, 178, 371
\bibitem[\protect\citeauthoryear{Pesce, Capuzzo-Dolcetta \& Vietri}{1989}]{pesce92} Pesce, E., Capuzzo-Dolcetta R., Vietri, M. 1992, MNRAS, 256, 368
\bibitem[{{Peuten} {et~al.}(2016)}]{Peuten16} Peuten, M., Zocchi, A., Gieles, M., Gualandris, A., H\'{e}nault-Brunet, V. 2016, MNRAS, 462, 2333

\bibitem[\protect\citeauthoryear{Repetto et al.}{2012}]{repetto12} Repetto S., Davies, M.B., Sigurdsson, S. 2012, MNRAS, 425, 2799
\bibitem[\protect\citeauthoryear{Rodriguez et al.}{2015}]{rodriguez15} Rodriguez, C. L., Morscher, M., Pattabiraman, B., Chatterjee, S., Haster, C.-J., and Rasio, F. A. 2015, Physical Review Letters, 115, 051101
\bibitem[\protect\citeauthoryear{Rodriguez et al.}{2016}]{rodriguez16} Rodriguez, C. L., Chatterjee, S. and Rasio, F. A. 2016, Physical Review Letters, 93, 084029

\bibitem[\protect\citeauthoryear{Rozier et al.}{2019}]{Rozier19} Rozier, S., Fouvry, J.B, Breen, P.G., Varri, A.L., Pichon, C., Heggie, D. C., 2019, MNRAS, Submitted, arXiv:1902.09299


\bibitem[\protect\citeauthoryear{Spera, Mapelli \& Bressan}{2015}]{spera15} Spera, M., Mapelli, M., Bressan, A. 2015, MNRAS, 451, 4086

\bibitem[\protect\citeauthoryear{Spera, Mapelli \& Jeffries}{2016}]{spera16} Spera, M., Mapelli, M., Jeffries, R.D. 2016, MNRAS,460, 1

\bibitem[\protect\citeauthoryear{Spera \& Mapelli}{2017}]{spera17} Spera, M. \& Mapelli, M. 2017, MNRAS,470.4739


\bibitem[\protect\citeauthoryear{Tiongco et al.}{2017}]{Tiongco17} Tiongco M. A., Vesperini E., Varri A. L., 2017, MNRAS, 469, 683

\bibitem[\protect\citeauthoryear{Tremaine \& Weinberg}{1984}]{tremaine84} Tremaine S. D. \& Weinberg, M.D. 1984, MNRAS, 209, 729

\bibitem[{{Trenti} {et~al.}(2010)}]{Trenti10} Trenti, M., Vesperini, E., Pasquato, M. 2010, ApJ, 708, 1598
\bibitem[\protect\citeauthoryear{Trenti \& van der Marel}{2013}]{trenti13} Trenti M., van der Marel, R. 2013, MNRAS, 435, 3272
\bibitem[\protect\citeauthoryear{Valtonen \& Karttunen}{2006}]{valtonen06} Valtonen M., Karttunen H. 2006, The Three-Body Problem (Cambridge: Cambridge University Press)
\bibitem[\protect\citeauthoryear{Vicari, Capuzzo-Dolcetta \& Merritt}{2007}]{vicari07} Vicari A., Capuzzo-Dolcetta R., Merritt, D. 2007, ApJ, 662 797
\bibitem[{{Webb} \& {Vesperini}(2016)}]{Webb16} Webb, J.J. \& Vesperini, E. 2016, MNRAS, 463, 2383
\bibitem[\protect\citeauthoryear{Webb et al.}{2018}]{webb18} Webb J. J., Leigh N. W. C., Singh A., Ford K. E. S., McKernan B., Bellovary J. 2018, MNRAS, 474, 3835 

\bibitem[\protect\citeauthoryear{Weinberg}{1986}]{Weinberg86} Weinberg, M. D. 1986, ApJ, 300, 93


\end{thebibliography}
\end{document}